\documentclass[sigconf]{acmart}
\acmConference[SE 2030]{International Workshop on Software Engineering in 2030}{November 2024}{Puerto Galinàs (Brazil)}
\AtBeginDocument{%
  \providecommand\BibTeX{{%
    \normalfont B\kern-0.5em{\scshape i\kern-0.25em b}\kern-0.8em\TeX}}}

\setcopyright{acmlicensed}
\copyrightyear{2024}
\acmYear{2024}
\acmDOI{XXXXXXX.XXXXXXX}

\usepackage{import}
\usepackage{multirow}
\usepackage{graphicx}

\acmJournal{JACM}
\acmVolume{37}
\acmNumber{4}
\acmArticle{111}
\acmMonth{8}

\begin{document}

\title{Integrating Sustainability Concerns into Agile Software Development Process}

\author{Shola Oyedeji}
\authornote{All authors contributed equally to this research.}
\orcid{1234-5678-9012}
\affiliation{%
  \institution{LUT University}
  \city{Lappeenranta}
  \country{Finland}
  \postcode{53850}
}
\email{shola.oyedeji@lut.fi}
\author{Ruzanna Chitchyan}
\affiliation{%
  \institution{University of Bristol}
  \city{Bristol}
  \country{UK}}
\email{r.chitchyan@bristol.ac.uk}
\author{Mikhail Ola Adisa}
\affiliation{%
  \institution{LUT University}
  \city{Lappeenranta}
  \country{Finland}
  \postcode{53850}
}
\email{mikhail.adisa@lut.fi}

\author{Hatef Shamshiri}
\affiliation{%
  \institution{LUT University}
  \city{Lappeenranta}
  \country{Finland}
  \postcode{53850}
}
\email{hatef.shamshiri@lut.fi}
\renewcommand{\shortauthors}{Trovato and Tobin, et al.}

\begin{abstract}
Software has the potential to be a key driver in fostering sustainability. Despite this potential, it is not clear if and how the software industry integrates consideration of sustainability into its common software development processes.
This research starts by investigating the current state of sustainability consideration within the software engineering industry through a survey. The results highlight a lack of progress in practically integrating sustainability considerations into software development activities.
To address this gap, a case study with an industry partner is conducted to demonstrate how sustainability concerns and effects can be integrated into agile software development. The findings of this case study demonstrate practical approaches to integrating sustainability into software development practices. 
Reflecting on the findings from the survey and the case study, we note some insights on scaling up the adoption of sustainability consideration into the daily practice of agile software development.

\end{abstract}



\received{6 April 2024}
\received[accepted]{9 May 2024}

\maketitle

\section{Introduction}\label{sec:intro}

Software systems play a pivotal role in all human activities [4, 15]. On the one hand, software systems are crucial in driving economic transformation and digitalization across various sectors - presenting opportunities to mitigate environmental impacts, enhance social equity, and improve individual well-being. On the other hand, these very systems can undermine sustainability via, for example, resource depletion for hardware and software production needs, related carbon emissions, and biodiversity degradation, as well as exacerbating societal inequalities due to 'baking'  biases and exclusions into societal infrastructures \cite{su10103471} \cite{Sreedhar}.  

To illustrate the magnitude of this prospective dual impact in terms of global CO$_{2}$ equivalent emissions, it is reported that at present the (rapidly growing) impact from ICT itself amounts to roughly 3\% \citep{FREITAG2021100340} and is estimated to exceed 14\% of the 2016-levels by 2040 \cite{Belkhir2018}. At the same time, ICT is claimed to be able to reduce the global footprint of other sectors by up to 15\% \cite{EricssonReport-ReduceFootprint}. It is not a surprise, then, that the software industry needs to concertedly address the sustainability issues of their products. 

The first calls to the Software Engineering (SE) profession to urgently integrate sustainability consideration into their daily practices to prevent the further growth of software's negative environmental impacts and foster its positive change have been made nearly 10 years ago \cite{becker2015sustainability,chitchyan2016sustainability}. But \textit{has the profession changed its practice?} And, if not, \textit{how could the current practice be adapted to account for sustainability in SE?}  
These are the two key questions addressed in the present paper. 
% 
% This paper recognizes the importance of addressing this sustainability engineering gap and takes proactive steps to investigate how sustainability considerations can be embedded into agile software development practices (which is the most prominent Software Engineering practice today). Such integration could ensure that SE remains efficient as a discipline and contributes positively to the sustainability goals of other domains. 

The paper starts by outlining the background of the research on SE and sustainability in section \ref{sec:background}. Section \ref{sec:methodology} details the study design, providing a perspective on how the notion of sustainability is treated in this study. The results for the current state of sustainability practice in SE among software development practitioners are in section \ref{sec:motivation}. Driven by the observed lack of progress in SE on practically integrating sustainability consideration into software development activities, we undertake a case study with an industry partner (software development company) to illustrate how their practice can be adapted to develop software systems with sustainability in mind as discussed in section \ref{sec:approach}. Section \ref{sec:roadmap} covers discussion and a roadmap for sustainability integration in agile software development processes. Concluding remarks and future work are in section \ref{sec:conclusion}. The key contributions of this research are in:
\begin{itemize}
    \item Providing an overview on the current state of integrating sustainability into SE practice and highlighting the lack of notable progress on this state compared to that of 10 years ago (see section \ref{sec:motivation}).
    \item Demonstrating a preliminary practicable approach towards integrating sustainability into agile software engineering practice using an ongoing case study with an industry partner (see section \ref{sec:approach}).
    \item Outlining the roadmap of research to integrating sustainability into SE practice for 2030 (see section \ref{sec:roadmap}).
\end{itemize}
\section{Background: Sustainability in Software Engineering}\label{sec:background}
Studies of what sustainability is for the SE domain have been wide and varied in recent years. For instance, Penzenstadler \cite{gustavsson2020blinded} describes sustainability as a comprehensive and holistic process of balancing environmental, social, and economic dimensions to achieve and maintain a desired state. Sedano \cite{sedano2016sustainable} describes sustainable software development as the capability and tendency of a software development team to address the adverse impacts of significant disruptions. More generally, sustainability in SE is seen from 2 viewpoints: \textit{sustainable software} and \textit{sustainability by software}. A {Sustainable software} is that whose production, deployment, and usage do not adversely affect the economy, society, or the environment \cite{schmidt2016sustainability,dick2013green,kern2015labelling}.
On the other hand, \textit{sustainability by software} is focused on the use of software to accomplish sustainability-focused goals \cite{condori2019towards}. 

In this research (following the conceptual framework set out by the Sustainability Awareness Framework \cite{duboc2019we,betz2024lessons}), the notion of sustainability is discussed as a systemic concern comprised of the following dimensions \cite{becker2015sustainability}:
\begin{itemize}
    \item \textit{Environmental}, which is focused on the use and maintenance of natural resources.
    \item \textit{Social}, focused on factors that affect the interaction between groups of people or communities, such as trust and equality.
    \item \textit{Individual}, which concerns the well-being of individuals and their equal access to services.
    \item \textit{Technical}, which addresses the ease of the software system's change, maintenance, and evolution. 
    \item \textit{Economic}, which focuses on maintaining the financial and capital assets of the businesses that develop and/or operate the software solutions.
\end{itemize}
Moreover, the impacts an ICT solution makes on the sustainability of its in-situ socio-economic environment are expected to materialize through time in 3 categories:
\begin{itemize}
    \item Immediate (direct) effects resulting from the development of the software systems (e.g., due to use of materials and energy). 
    \item Enabling (indirect) effects: effects resulting from the use of software systems.
   \item Structural (systemic) effects: Based on the long-term usage of the software systems which shifts (societal) practices, typically taking years to manifest visibly.
\end{itemize}

Several researchers \cite{duboc2019we,chitchyan2022theory,betz2024lessons,chitchyan2024can} argue that when software designs are informed with sustainability requirements, the resulting software and their socio-technical systems would be conducive to sustainability. Studies on considering sustainability specifically when carrying out the agile software development process \cite{barroca2018sustaining,gregory2016challenges,karita2021software,kasurinen2017concerns,bambazek2023} have recently been invigorated. Yet, \textit{the integration of sustainability concerns into the agile development process remains poorly aligned and piece-meal}. For instance,  while methods have been suggested for eliciting sustainability-related requirements \cite{duboc2019we,chitchyan2022theory}, these have been applied separately from the 'normal' agile user story creation process. Thus, it has not been clear how to integrate sustainability requirements into agile software delivery practice. 

Venters et al. \cite{venters2023sustainable} point out the lack of metrics to evaluate the sustainability of software across various domains and this applies to agile development as well. In summary, as noted by Adisa et al. \cite{adisa2022software}, the delivery of sustainable software will remain impeded until developers are enabled with tools and methods for sustainable design practices which are \textit{integrated into the software development process}. Yet, such practical methods and tools are still in short supply \cite{duboc2019we, chitchyan2016sustainability}.

\section{Methodology}\label{sec:methodology}
Given that this paper set out to identify if and how the calls to integrate sustainability into software engineering practice have been taken up by the current SE industry, the following two key research questions were defined:
\begin{enumerate}
    \item \textbf{RQ1}: What is the current state of sustainability considerations within SE practice? 
    \item \textbf{RQ2}: How (i.e., through what activities and at which steps) does/would the agile software development process integrate sustainability? 
\end{enumerate}

To address the first question, we designed a `light-touch' (5 minutes long) survey approach to elicit a general picture of the sustainability considerations in SE practice. For the second question, a more detailed consideration of the processes that integrate sustainability into SE practice was carried out through a case study. 

\subsection{Survey Design}\label{sec:surveyDesign}

The survey sought to collate data on general software development methodologies employed by software development practitioners in the industry and the sustainability factors taken into account during the software development process. 
The \textbf{Survey Questions} were designed as multiple-choice, open-ended, and Likert scale questions to gather viewpoints on software development methodologies and sustainability considerations. Topics addressed in the survey are shown in Table \ref{tab:review}. 
\begin{table*}[]
\caption{Software Development Practitioner's Survey Responses}
\label{tab:review}
\resizebox{\textwidth}{!}{%
\begin{tabular}{l|l|l|llll}
\cline{1-5}
\multicolumn{1}{|l|}{\textbf{Years of Experience}} &
  \textbf{0-3} &
  \textbf{3-5} &
  \multicolumn{1}{l|}{\textbf{5-10}} &
  \multicolumn{1}{l|}{\textbf{Over 10}} &
  \multicolumn{2}{l}{\multirow{4}{*}{}} \\ \cline{1-5}
 &
  18 (10\%) &
  64 (35\%) &
  \multicolumn{1}{l|}{83 (47\%)} &
  \multicolumn{1}{l|}{13 (7\%)} &
  \multicolumn{2}{l}{} \\ \cline{1-5}
\multicolumn{1}{|l|}{\textbf{What methodologies does your team use?}} &
  \textbf{Waterfall} &
  \textbf{Agile} &
  \multicolumn{2}{c}{\multirow{2}{*}{\textbf{}}} &
  \multicolumn{2}{l}{} \\ \cline{1-3}
 &
  3 (1\%) &
  116 (99\%) &
  \multicolumn{2}{c}{} &
  \multicolumn{2}{l}{} \\ \cline{1-6}
\multicolumn{1}{|l|}{\textbf{How frequently does you team   consider sustainability activities in SDL?}} &
  \textbf{Never} &
  \textbf{Rarely} &
  \multicolumn{1}{l|}{\textbf{Occacionally}} &
  \multicolumn{1}{l|}{\textbf{Often}} &
  \multicolumn{1}{l|}{\textbf{Always}} &
  \multirow{2}{*}{} \\ \cline{1-6}
 &
  89 (50\%) &
  79  (45 \%) &
  \multicolumn{1}{l|}{9 (5\%)} &
  \multicolumn{1}{l|}{0} &
  \multicolumn{1}{l|}{0} &
   \\ \hline
\multicolumn{1}{|l|}{\textbf{Which sustainability dimensions do you consider during development?}} &
  \textbf{None} &
  \textbf{Economic} &
  \multicolumn{1}{l|}
  {\textbf{Technical}} &
  \multicolumn{1}{l|}{\textbf{Environmental}} &
  \multicolumn{1}{l|}{\textbf{Individual}} &
  \multicolumn{1}{l|}{\textbf{Social}} \\ \hline
 &
  78 (40\%) &
  55 (28\%) &
  \multicolumn{1}{l|}{51 (26\%)} &
  \multicolumn{1}{l|}{12 (6\%)} &
  \multicolumn{1}{l|}{0\%} &
  \multicolumn{1}{l|}{0\%} \\ \hline
\multicolumn{1}{|l|}{\textbf{How does the chosen development methodology impact sustainability?}} &
  \textbf{I do not know} &
  \textbf{No Influence} &
  \multicolumn{1}{l|}{\textbf{Positively}} &
  \multicolumn{1}{l|}{\textbf{Negatively}} &
  \multicolumn{2}{l}{\multirow{3}{*}{}} \\ \cline{1-5}
 &
  106 (60\%) &
  44 (25\%) &
  \multicolumn{1}{l|}{22 (12\%)} &
  \multicolumn{1}{l|}{6 (3\%)} &
  \multicolumn{2}{l}{} \\ \cline{1-5}
\multicolumn{1}{|l|}{\textbf{Does your   company have explicit policies for sustainability in software development?}} &
  \textbf{No} &
  \textbf{Not sure} &
  \multicolumn{1}{l|}{\textbf{Yes}} &
   &
  \multicolumn{2}{l}{} \\ \cline{1-4}
 &
  135 (76\%) &
  43 (24\%) &
  \multicolumn{1}{l|}{0} &
  \multicolumn{3}{l}{\multirow{3}{*}{}} \\ \cline{1-4}
\multicolumn{1}{|l|}{\textbf{Have you   received training on sustainability for software development}} &
  \textbf{No} &
  \textbf{Yes, basic} &
  \multicolumn{1}{l|}{\textbf{Yes, extensive}} &
  \multicolumn{3}{l}{} \\ \cline{1-4}
 &
  171 (96\%) &
  7 (4\%) &
  \multicolumn{1}{l|}{0} &
  \multicolumn{3}{l}{} \\ \cline{2-4}
\end{tabular}%
}
\end{table*}

\textbf{Recruitment of survey participants} was carried out through a range of methods, such as disseminating the survey through industry associations, internet platforms for software experts, and professional networks. To attract participants, the survey link had to be shared with a brief description of the study's goals and guarantees of anonymity and confidentiality.
    
\textbf{Data Analysis:} After the data was gathered, Webropol and Excel were used to combine and analyze the responses in order to get important insights. 

The survey followed \textbf{ethical guidelines set out by the University of Bristol}, guaranteeing minimal data collection and use of the data for research purposes only, as well as participant anonymity and confidentiality. 

\subsection{Case Study Design} \label{sec:caseStudyDesign}
A collaborative case study was used to explore avenues to support software development practitioners with the integration of sustainability considerations into agile software development (specifically the Scrum Framework). The process comprised partnership formation and data collection and analysis.

The \textbf{partnership formation} started when a product manager from a software development company contacted the researchers to collaborate on supporting their software development team with investigating ways to incorporate sustainability concerns and impacts into the company's agile software development processes. Table \ref{tab:practitioners} details the practitioners (i.e., the Agile Development team) who were then involved with the case study.
\begin{table}[H]
\caption{Software development practitioners from the company}
\label{tab:practitioners}
\resizebox{\columnwidth}{!}{%
\begin{tabular}{|l|l|l|}
\hline
No & Job Title                    & Years of Experience \\ \hline
1  & Scrum Master                 & 9                   \\ \hline
2  & Product Owner                & 6                   \\ \hline
3  & Programmer                   & 6                   \\ \hline
4  & Software Developer/Architect & 6                   \\ \hline
5  & UX Specialist                & 5                   \\ \hline
6  & IT Manager                   & 4                   \\ \hline
\end{tabular}%
}
\end{table}
\textbf{Data collection} was carried out through workshop sessions conducted by the company, where the Agile Development team shared their experiences. Information on their development processes, practices, and sustainability considerations was shared. 

The researchers used \textbf{thematic analysis} to study the collected data and discovered practice patterns and insights related to sustainability integration in agile software context. Based on this, the researchers and the agile team co-designed processes that integrated sustainability considerations into the company's software development processes. 
    
As a result, the Sustainability Agile Requirement Toolbox (SART) was created which included sustainability dimensions and effects for every sprint item in the Scrum framework’s product backlog, accompanied by a sustainability sprint retrospective template. 
    
This case study is still ongoing, undertaking continuous data collection and collaboration with the Agile Development team for refinement of strategies, assessment of implementation outcomes, and iterative improvement of sustainability practices within agile development processes.

\subsection{Threats to Validity}\label{sec:threat-validity}
Below we note a number of issues of relevance to the validity of this study.
% Both the survey and the case study were constructed using  Existing frameworks, literature studies, and experts (software development practitioners and researchers) input were used to operationalize study concepts including software development methods, agile methodologies, and sustainability integration in software development. Survey instruments and a case study from a company involving an agile software development team with more than three years of experience were used to inform our research investigation.

With respect to the survey component:
\begin{itemize}
    \item The survey was designed to elicit the details on the current practice within the software industry. However, we also asked about the notion of `sustainability' with clear expectation that this construct can be interpreted in many ways. This was done to observe the perception of the respondents on their engagement with sustainability as a topic.  Once the respondents answered if they did/did not use sustainability in their SE practice, they were also given examples and asked to choose which areas of sustainability they did/did not engage with. The areas of sustainability are based on the SuSAF model \cite{duboc2019we}. For instance, asking if the respondents addressed the environmental dimension was accompanied by examples of energy consumption and carbon emissions, etc. Thus, we observe both the general perception on sustainability engagement and the more specific focus on engagement with dimensions with examples. Nevertheless, it is still possible that the interpretation of sustainability and its various dimensions differs somewhat across the respondents. 
    \item For the survey distribution, given that the survey was distributed among software development companies in Europe and a professional network initiated by European software practitioners, the representation of respondents might be dominated by practitioners from Europe. This can impact the global generalization of the results. However, the findings present, at the very least, the current European view.  
\end{itemize}

With respect to the case study component:
\begin{itemize}
    \item The choice of the partner for this case study was driven by the partner's interest itself. The researchers did not set out any objective selection criteria. Thus, we make no claims on the representatives of this company or their context. Given that the knowledge, motivation, and buy-in of the team members impacts the ease and speed of the process change, it is likely that teams with other context will have different barriers to adopting the suggested process changes.  Nevertheless, given that the key issue researched with the case study is the feasibility of the process change itself, we consider this to be an acceptable way of case study choice. 
    \item The case study itself addresses development of a commercially sensitive product. Thus, while the researchers were given access to some sprint reviews and some materials, the content of the user stories (e.g., in the sprint review templates) was heavily reduced and modified. This, while richness of the case study context, does not impede drawing the conclusions on the methods success and utility. 
\end{itemize}

\section{Survey: Sustainability in Software Engineering Practice}\label{sec:motivation}
This section details the results of a survey designed to ascertain the current state of sustainability consideration among software development practitioners (to address RQ1). The survey was conducted between 01-01-2023 and 18-03-2024 and received 178 responses, which were summarized in Table \ref{tab:review}.

The survey respondents were software development practitioners with significant experience in SE practice (54\% over 5 years of experience, 36\% with 3 to 5 years of experience, and only 10\% with less than 3 years of experience).  The majority of them work in the core ICT sector (78\% of respondents), though other sectors are also represented, including Transportation (6\%), Construction (6\%), Finance and Insurance (6\%), and Electricity and Gas (4\%).

For software development methodologies, 99\% of respondents reported using Agile and only 1\% (3 respondents) used 
% which is more common than DevOps (13\%), Lean (3\%), 
 Waterfall. 
 % Out of the Agile methodology used in software development processes, scrum holds a 71\% share, while Kanban holds a 29\% share. 
 Nevertheless, despite the wide use of Agile approaches, the respondents do not sufficiently incorporate sustainability concerns and impacts analysis into software development: only 5\% of respondents occasionally consider sustainability impacts on the economy and environment, whilst 95\% reported rarely or never considering sustainability. Furthermore, when asked which sustainability dimensions were taken into account when developing software, a substantial percentage of respondents (40\%) did not consider any sustainability dimensions, followed by 28\% who mentioned the economic dimension, 26\% for the technical dimension, and 6\% the environmental dimension. The social and individual dimensions were not considered by software practitioners at all.

Furthermore, regarding how selected software development approaches affected sustainability, the majority of respondents (60\%) reported that they had no idea, 25\% reported it does not affect sustainability,  12\% reported it has a positive influence on sustainability, and 3\% reported it can negatively affect sustainability. This lack of awareness suggests a disconnection between software development methodologies and their potential influence on sustainability. Additionally, regarding organizational policies or guidelines for integrating sustainability into development processes, most of the respondents (76\%) said that they had no policy in place, with the rest (24\%) being unsure whether such guidelines were used.

Regarding education and training on sustainability among software development practitioners, the results indicate that only 4\% of respondents had received basic training with an emphasis on green coding, energy efficiency, the Sustainability Awareness Framework (SusAF), and reading the Karlskrona Manifesto for Sustainability Design. However, the majority of respondents (96\%) had not received any training or education on sustainability and its relation to software development. This demonstrates the need for more thorough education and sustainability-related training in software engineering for current and future software development practitioners. 

The overall findings of the survey indicate a concerning gap in awareness and integration of sustainability in the current state of software engineering and development practices. Despite the widespread adoption of Agile methodologies, sustainability effects and impacts are not well understood or integrated into software development processes.
\section{Case Study: Sustainability in Agile Software Engineering}\label{sec:approach}
\begin{figure*}[]
    \centering
\includegraphics[width=0.95\linewidth]{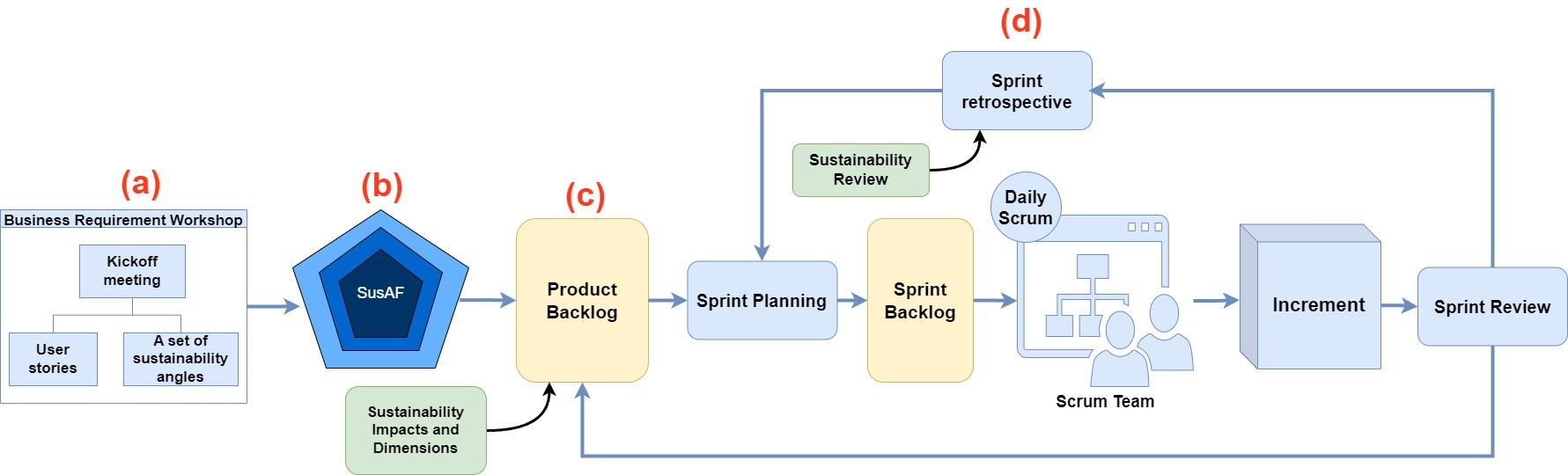}
    \caption{Integrating Sustainability into Agile Process (Scrum)}
    \label{fig:agileWorkshop}
\end{figure*}

The collaborative case study to address RQ2 on engineering activities and practices is outlined in Figure \ref{fig:agileWorkshop} with the study participants listed in Table \ref{tab:practitioners}.
 The study commenced with a review of the company's software development processes and practices at a \textbf{business requirements workshop} (see Fig.\ref{fig:agileWorkshop} (a)) using a planned software as a guiding example. 
 To maintain anonymity, the case study software will henceforth be called ``Labor Hire"; it is an online marketplace connecting labor service providers with service seekers, including companies and households.
 The key activities of the reported process are briefly summarized below:
\begin{itemize}
    \item Requirements were gathered through user stories involving customers/users of the software products and the agile team (product owner, developers, UX expert).
    \item User and customer interviews were carried out, involving direct communication between the design and development teams to better clarify user needs and ensure alignment during requirement elicitation.
    \item Using a template, each team member took notes during customer or user interviews to cross-reference and delve deeper into customer problems and needs for the software products and services.
    % \item When there are limitations that make it difficult to communicate with real users, personas are occasionally employed.
    \item Software product and service development was carried out using the Scrum agile framework through several sprints.
\end{itemize}
The company wished to integrate sustainability concerns into their software development process but lacked knowledge of methods and experience in this. Considering the outlined user-centered requirements elicitation process and the lack of understanding of sustainability concerns, the researchers proposed the Sustainability Awareness Framework (SusAF) \cite{SuSAFworkbook,duboc2019we} as a tool well-suited to the given context for sustainability considerations. SusAF requires both engaged conversations with stakeholders for the identification of sustainability impacts of the developed software systems (across social, individual, environmental, economic, and technical dimensions as well as for direct, indirect, and systemic effects) and provides a set of questions that facilitate the sustainability-related impact elicitation. 

\begin{figure}[h]
    \centering
\includegraphics[width=0.95\linewidth]{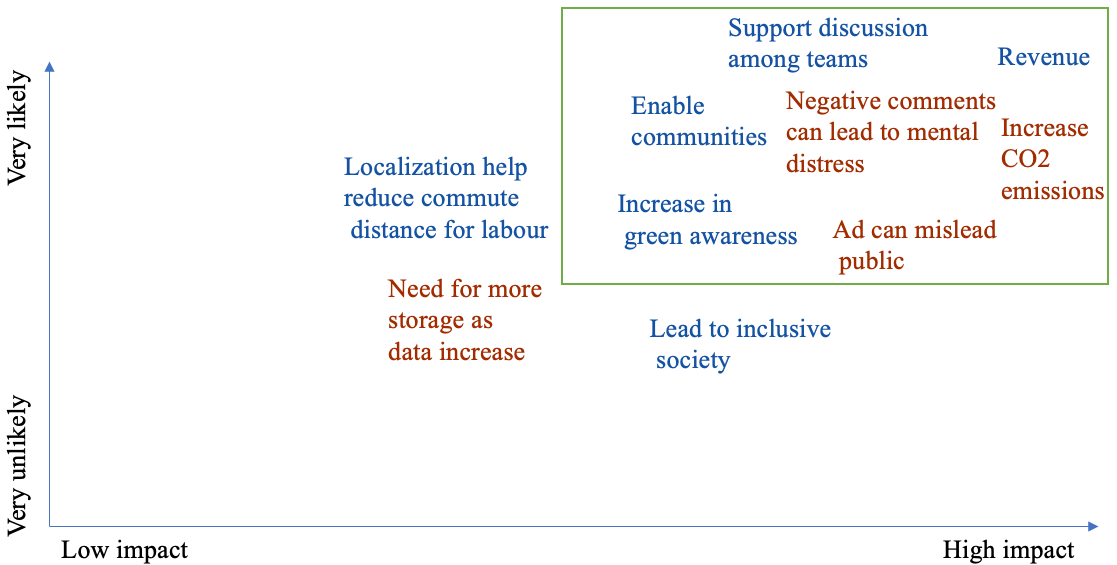}
    \caption{Prioritize – Likelihood and Level of Impacts for the Labor Hire software }
\label{fig:Prioritize-Impacts}
\end{figure}

The suggested use of SusAF was deemed relevant by the agile team, and a \textbf{SusAF application workshop} was arranged (see Fig. \ref{fig:agileWorkshop} (b)) to use the framework to elicit the potential sustainability impacts of Labor Hire. 
The workshop participants identified the sustainability impacts of the Labor Hire software (illustrated in Figure \ref{fig:Prioritize-Impacts}) and prioritized these based on how likely each of these impacts was. The highly likely and strongly impactful effects are shown within the square box of Figure \ref{fig:Prioritize-Impacts}, and those with lower impact and likeliness are shown outside the square box.  

\begin{figure}[h]
    \centering
\includegraphics[width=0.95\linewidth]{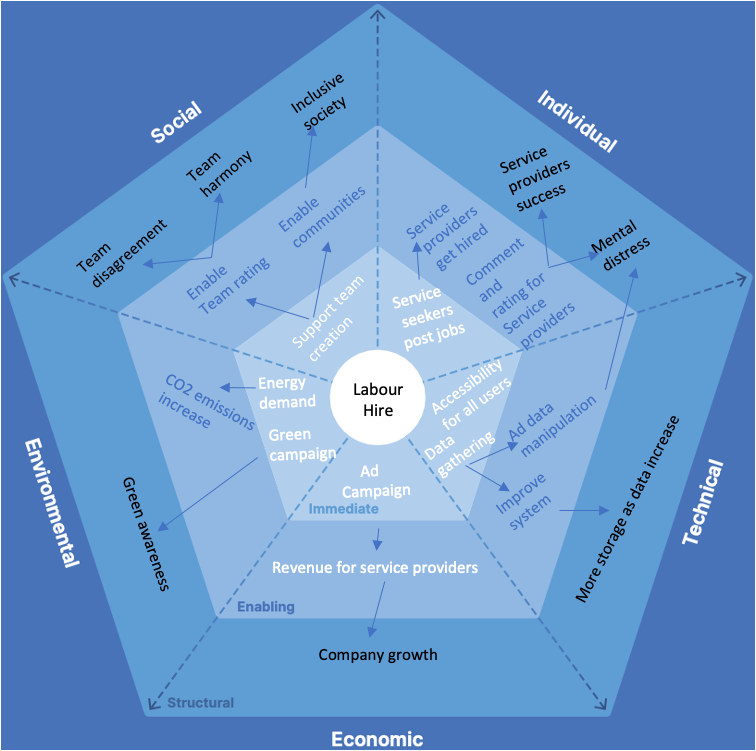}
    \caption{Labor Hire Chain of Sustainability Effects}
    \label{fig:Chain-of-effects}
\end{figure}

The SusAF tool also suggests utilizing the sustainability awareness diagram (SuSAD) - an adapted radar chart used to facilitate discussions on \textit{chains of effects} (i.e., how impact on one aspect could evolve over time and cause other impacts). The SuSAD for the sustainability chain of effects for the Labor Hire software is shown in Figure  \ref{fig:Chain-of-effects}. This figure, drafted as part of an in-depth discussion among the study participants, facilitated the examination of the software's chain of effects and prompted the agile team to seek ways of integrating the impacts  and effects identified through SusAF into their their Scrum process. 

\begin{figure*}[h]
    \centering
\includegraphics[width=0.95\linewidth]{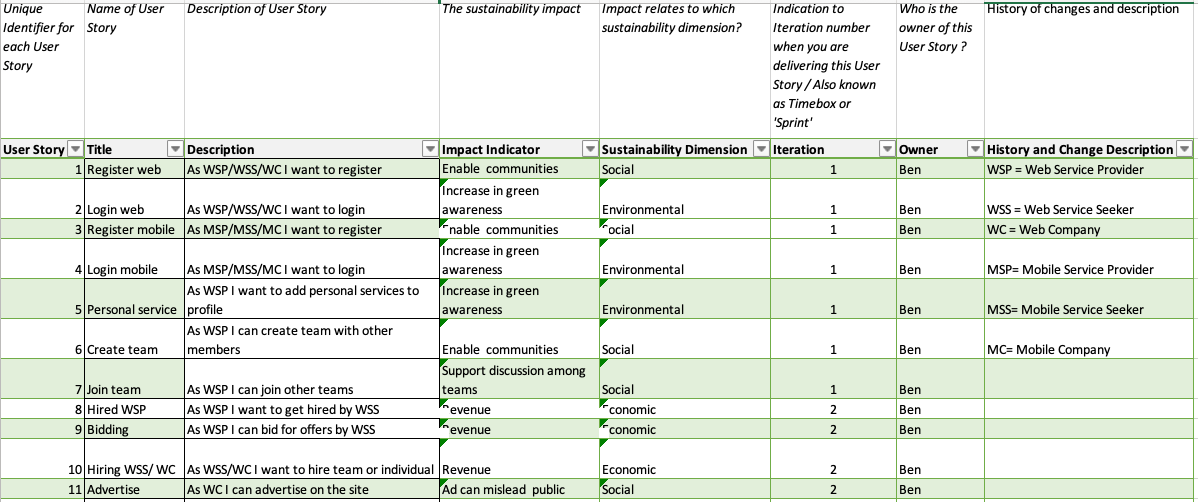}
    \caption{SART product backlog for the Labor Hire software system}
    \label{fig:SART-product-backlog}
\end{figure*}

To this end, the Scrum's product backlog was collaboratively re-designed with the \textbf{inclusion of sustainability dimensions} and effects for every item in the \textbf{product backlog} (see Fig.\ref{fig:agileWorkshop} (c)), and a \textbf{sustainability sprint retrospective} template was used to review the sustainability impacts of each sprint (see Fig.\ref{fig:agileWorkshop} (d)). Figure \ref{fig:SART-product-backlog} illustrates an example of the adapted product backlog for the Minimum Viable Product of the Labor Hire system.
% Considering the identified sustainability impacts and effects in this manner assisted the scrum-based agile teams in taking sustainability into account while making design decisions for software development.  
%

Through these adaptations, software practitioners linked each item in the product backlog to a sustainability impact and dimension. This facilitated the agile team in addressing sustainability concerns throughout each sprint as they designed and developed the Labor Hire software. 

Every sprint concluded with a sustainability review that was part of the sprint retrospective process. An \textit{example extract from the sustainability retrospective }of the second sprint in this case study is noted below:
\begin{enumerate}
    \item \textbf{What helped our sustainability design decision?}
\begin{itemize}
    \item The short sprint was helpful during this experimental development with sustainability as a key factor.
    \item Having one developer play the role of sustainability rep helped to keep us on track as we made design decisions during this sprint.
\end{itemize}

\item \textbf{What were the obstacles and challenges?}
\begin{itemize}
    \item It was challenging to use Jira and the Excel template at the same time during this sprint.
    \item Most of us lack sustainability knowledge and have to rely on the SusAF results, templates, and the researchers' knowledge.
\end{itemize}

\item \textbf{What can we do differently to improve the overall sprint process?}
\begin{itemize}
    \item Incorporate the sustainability impacts into the sprint backlog.
    \item Establish some metrics to measure the sustainability of each sprint outcome.
\end{itemize}

\item \textbf{What should we incorporate in the next sprint? }
\begin{itemize}
    \item Add sustainability impacts the sprint backlog. 
    \item The researcher should join our sprint planning as a sustainability stakeholder.
\end{itemize}

\end{enumerate}
This review highlights the significance of having a dedicated sustainability representative within the agile development team for sustainability to be effectively integrated into software development. The agile team admitted that throughout the sprint, they ran across problems as a result of their lack of experience with sustainability, which they were able to resolve by using the SusAF results and backlog as well as the retrospective templates. 

Throughout each sprint, the agile team remained committed to incorporating sustainability considerations into the design and development of the Labor Hire software application. For example, some of the sustainability-focused features integrated into Labor Hire include a green awareness campaign module to educate users (both service providers and seekers) about sustainability; the agile development team designed the Labor Hire software system with accessibility in mind, ensuring usability for individuals with diverse physical and visual abilities. In addition, they implemented a conflict resolution mechanism within the Labor Hire application to mitigate disputes among service provider teams formed by service seekers for projects. To further enhance user experience, content filtering tools were added to detect and eliminate hateful remarks from user accounts.

While this is an ongoing case study, some of the comments that the team provided when asked about what has changed in their practice so far with use of SusAF and adapted Scrum process for 5 sprints noted that:
\begin{itemize}
    \item Throughout the sprints, the team now holds conversations about their mental and physical well-being
    \item The sustainability footprint of the software product is now evaluated at the end of each sprint using the sustainability impacts and effects found using SusAF
    \item To make comparisons with other providers easier, the team proactively asked their cloud service providers for information on their CO2 emissions during deployment
    \item The team has committed to ongoing sustainability dialogues within their company, enabling them to make more informed decisions. For instance, while transitioning to more sustainable cloud service providers might incur additional costs, they are exploring options to influence their existing partners and suppliers to reduce environmental impact
    \item Emphasizing inclusivity and community engagement, the team has prioritized accessibility in their product development endeavors
    \item The company has now started initiatives to investigate green coding techniques and create teaching plans for their junior developers
    \item The team is developing strategies to reduce their development-related carbon emissions; this has become a critical component of their software development efforts.
\end{itemize}

In summary, our collaborative exploration of integrating sustainability into the agile software engineering process of the Labor Hire team has led to transformative changes within the process used by the team, the team's mental model on sustainability, and the development company's activities. As we co-developed ways to integrate sustainability considerations and effects into the team's Scrum process, it also progressively advanced the team's understanding of sustainability in software development. 
\section{Discussing the Roadmap to 2030}\label{sec:roadmap}
Software systems play a key role in shaping society through digitization, with profound sustainability impacts on users, society, the environment, and the economy. Software systems have the potential to substantially reduce negative impacts caused by human activities within a of a vast variety of domains. Yet, this can only occur if sustainability considerations are  integrated into the design, development and operation of these systems themselves. What lessons have we learned from the above discussed survey and case study of Labor Hire, and what key milestones must be passed for us to arrive at a sustainable ICT ecosystem for 2030?

\subsection{Reflection on Findings So Far}
Turning back to our survey study, presented in section \ref{sec:motivation}, we must unfortunately conclude that the software engineering profession has not yet assumed the responsibility of integrating sustainability into ICT products and services. As previously noted, 95\% of our survey respondents consider sustainability only rarely or not at all. In the rare cases when sustainability is considered, only in 6\% of those did the respondents note the relevance of environmental concerns, while social and individual sustainability concerns seem to be missing from the worldview of software engineering practitioners altogether. Thus, \textit{the present state of practice needs urgent and dramatic change if the software industry is to be sustainable in the medium term}.

Fortunately, as illustrated by the (ongoing) Labor Hire case study discussed in section \ref{sec:approach}, \textit{all key components necessary for such a dramatic change towards integrating sustainability into software engineering practice are already present}. More specifically:
\begin{itemize}
    \item the SuSAF framework supports the familiarization of software practitioners with the notions of sustainability (i.e., its 5 dimensions, orders of effects, chains of effects and impact magnitude), as well as
    \item provides a simple set of questions that foster the initial identification of sustainability requirements and effects.
    \item The adapted product backlog templates and sprint review backlogs allow for each software and hardware issue in the backlog to be considered for their impact on sustainability.
\end{itemize}

It is even more encouraging to note that \textit{the agile development process (the most commonly used process of the present day) does not require any foundational changes to support accountability for sustainability-related practices.} The Labor Hire team was able to start practicing a sustainability-supportive software engineering process after only 2 workshops (i.e. after spending about 3 hours) with the researchers. This demonstrates that an agile development team starting with no previous knowledge or expertise save that of the agile process can integrate sustainability considerations and expand their own software engineering practices quite immediately.

We also note that the few simple \textit{changes we co-designed into the team's Scrum process resulted in an astonishing impact on the mindset of the agile development team} of the Labor Hire system:
\begin{itemize}
    \item The team began by adjusting their own development practice so as to improve the sustainability impact of their product.
    \item The team has now started to also focus on the individual and societal well-being of the team members themselves (e.g., discussing the team's well-being as part of each sprint review, as well as planning for the training and up-skilling of the younger developers).
    \item Moreover, the team is progressing towards wider company-scale reviews of their suppliers (e.g., cloud service providers) with the aim of integrating sustainability holistically into their organization's wider ecosystem.
\end{itemize}

\subsection{Looking Forward to 2030}
Agile software development is the predominant approach to developing software systems today, and even though technological developments (such as code generation supported with AI) will likely change the toolbox used by the developers, we suggest that overall, the agile methodology will remain predominant in the medium future.  
Therefore, the roadmap is discussed with respect to integrating sustainability considerations into the agile software development process.
Although, as exemplified by the Labor Hire case study, agile teams can quickly integrate sustainability considerations into their daily practices, given the very low level of the current adoption of sustainability considerations in industry, notable improvement in the medium term will require \textit{the rapid scaling up of such integration}. Below we note three key directions for such up-scaling, each of which has a range of sub-activities.
\begin{itemize}
    \item \textbf{Engaging the Agile Manifesto}: To encourage the software development companies and practitioners at large to adopt sustainability considerations into their normal practices, the most effective change could be a change to the Agile Manifesto \cite{beck2001agile} itself. The Agile Manifesto is probably the only document that all current and upcoming generations of software engineers would take note, see or hear about at some point in their careers. It thus informs and guides the vast majority of developers' practice and could serve as a cornerstone of sustainability adoption. It is worth pointing out that some work in this direction has already commenced through the Agile for Sustainability \cite{Agile4Sustainability} and Agile Sustainability Manifestos \cite{AgileSustainabilityManifesto}. These can offer helpful starting points for motivating change. 
    \item \textbf{Standardization of Best Practices:} While integrating changes into individual projects is essential, it has to be done on a case-by-case basis. On the other hand, the provision of best practice standards can support the widespread adoption of sustainability-conducive changes. 
    The beginning of such standardization is already noted in some areas. For instance, AWS has released a framework on `Best practices for sustainability in the cloud' \cite{awsCloudSustainability} focusing on both improving the energy efficiency and carbon reduction of their services, and setting out guidelines for their service users, helping to integrate these considerations into their clients' practices. Similarly, the Green Software Foundation is working towards a collection of energy-efficient software development patterns \cite{GreenSoftwareFoundation}. Looking forward, a more coherent, software industry wide standardization would help to accelerate these processes. Example sub-activities relevant to this up-scaling and direction are related to industry engagement, evidence collection, evidence-based decision making, agreeing on metrics and measurement processes, development of tools that integrate the standards, developing domain-specific and cross-domain recommendations, etc. 
    \item \textbf{Education and Training:} It is imperative to enhance the current software engineering education offered by higher education institutions and training in companies to equip (future) software development practitioners with a deeper understanding of sustainability issues. This education should include both an introduction to the multi-dimensional and time-sensitive nature of sustainability and technical skills for related decision making and design/choice evaluation. As was illustrated by the Labor Hire team, once the team learned how to conceptualize sustainability (with the help of SusAF) and how to integrate it into their development process (through adapted backlog templates and spring reviews), they were empowered to take ownership of the sustainability impacts of the systems they develop. Example sub-activities relevant to this up-scaling direction are related to integrating systems thinking competencies into Computer Science curricula, learning to consider temporal impacts, accounting for values (held by both oneself and by others), working in cross-functional teams, etc.
\end{itemize}

In summary, we have discussed how a single team has been able to adopt sustainability considerations in their software development practice and noted 3 key directions for the rapid macro-upscaling of such adoption. Yet we also note that, because of the Labor Hire example, we have grounds to hope that agile teams can serve as catalysts for change within their organizations. Once a team is empowered to integrate sustainability into its practice, it can start driving efforts to evaluate and improve the sustainability of their software systems and that of the broader organizational ecosystem. In this way, engaged teams can fuel scaled-up change. 

\section{Conclusions}\label{sec:conclusion}
This research set out to identify whether the calls to integrate sustainability into the software engineering practice have been taken up by the current SE industry, and how this integration can be further fostered. To this end, we examined the current state of sustainability practices in software development and explored avenues for integrating sustainability into agile software development. Our research findings revealed a lack of awareness and action among software development practitioners, with only a few taking sustainability effects into account in their software development practices. 

The case study of Labor Hire that was undertaken with our industry partner demonstrated how Agile development teams can effectively identify and prioritize sustainability effects within Agile software development using SusAF. Integrating SusAF into the Scrum framework enabled the Agile development team to link sustainability effects to each item in the product backlog and conduct a sustainability retrospective at the end of each sprint.

Furthermore, the transformative shifts seen in the agile team's practices, mindset, and organizational activities demonstrate how sustainability integration can spark significant change in software development processes. A promising move towards more sustainable software engineering practices was shown by the Labor Hire system team's commitment to addressing sustainability concerns, fostering inclusivity, and engaging with stakeholders. Future work based on the ongoing case study will involve exploring with the agile development team the other components of the Scrum Framework (sprint planning and backlog, daily sprints, sprint review) that can facilitate better sustainability integration in agile software development and also report on the evolving software development practices within the company.

Looking ahead to 2030, we envision a future where sustainability is inherently integrated into agile software development approaches. Although, as discussed above, there has been some progress, industry stakeholders and educators must work together to ensure that sustainability practices are widely adopted and scaled up. We noted a few key directions for scaling up such adoption, including the integration of sustainability into the Agile Manifesto as a key document for current and future software development practitioners, the standardization of best practices for sustainable software development, and the expansion of software development practitioners' education and training to include sustainability-focused knowledge and implementation skills.

\begin{acks}

The authors express their gratitude to the respondents of the survey and the company collaborators who contributed to this study.
\end{acks}

\bibliographystyle{ACM-Reference-Format}
\bibliography{sample-base}

\appendix









\end{document}